# Analytical determination of bandgaps in photonic crystals

Sina Khorasani, Ali Adibi[*]
School of Electrical and Computer Engineering
Georgia Institute of Technology, Atlanta, GA 30332-0250, USA

**ABSTRACT**

A new approach for the analytical calculation of band structures in one-dimensional periodic media, such as photonic crystals and superlattices, is reported based on the recently reported differential transfer matrix method (DTMM). The media analyzed in this paper can have arbitrary profile of refractive index. A closed form dispersion equation is obtained and simplified under different assumptions. Several numerical cases are discussed.

## 1. INTRODUCTION

Photonic crystals have brought a possibility to fabricate novel optical devices, not generally foreseen previously. These structures are based on periodic variations of refractive index, with a length scale comparable to wavelength in the medium. Therefore, the propagation of electromagnetic waves in periodic structures has become the subject of extensive recent research.

One of the early attempts to study the wave propagation in multi-dimensional periodic media was made by Brillouin[1], for description of electron and phonon transport in electronic crystals. While the Brillouin approach is applicable to photonic crystals, it is still not suitable for many problems. In fact, the light propagation in layered stack of dielectrics could be described much easier through matrix approaches, as has been described by Yariv and Yeh[2,3] for the case of periodic stack of two alternating simple isotropic dielectrics.

Hall et al.[4] reported a scattering matrix formalism for the analysis of one-dimensional multi-layered periodic structures. Also Felbacq et al.[5] studied one-dimensional photonic crystals and deduced that infinite and semi-infinite photonic crystals have the same band structure, the only difference being the existence of surface states in the semi-infinite ones. They also considered the more general case of photonic crystal with arbitrary number of homogeneous layers and refractive indices in one period and found a universal curve for the location of band gaps in such Bragg mirrors. A recent study has discussed the more general case of one-dimensional photonic crystals of two alternating dielectric layers with arbitrary refractive index contrast and thicknesses[6] and presented approximate formulae for quick and accurate estimation of bandgap widths. Another report[7] has considered the Krönig-Penney photonic crystals obtained by conducting interfaces, and it has been shown that such structures as the electromagnetic dual of Krönig-Penney electronic crystals have exactly the same dispersion equation. In all of the above reports, the media considered have been layered structures of simple homogeneous dielectrics, with step-wise variations of refractive index.

The calculation of band structure of photonic crystals with arbitrary variation of refractive index is performed by plane wave expansion method[8]. However, this method is subject to poor convergence properties and requires the computation of eigenvalues of large matrices. In contrast, the recently reported finite element method[9] is capable of dealing with such problems in multi-dimensions, but using sparse matrices. Therefore, the efficiency and the accuracy are both improved. An approximate analytic method for the calculation of band structure of photonic crystals has also been reported[10], which is based on the asymptotic expansion of wave equation. But the method is generally good for qualitative, not quantitative, description of propagation phenomena.

---

[*] adibi@ece.gatech.edu; phone 1 404 385-2738 ; fax 1 404 894-4641

We have recently reported a new analytical tool for efficient and exact solution of one-dimensional wave equation in non-homogeneous media with arbitrary variations of refractive index[11] for TE modes, and its validity have been assessed through application to several examples. Extensions to TM modes and anisotropic non-homogeneous media have been also studied[12]. The method is based on the extension of transfer matrices into their differential form, and is referred to as Differential Transfer Matrix Method (DTMM).

Here, we report the application of the DTMM to the computation of the band structure of one-dimensional photonic crystals with arbitrary periodic variations of refractive index. We report a simple and general equation for the dispersion of Bloch waves and consider simplifications of that equation under even symmetry and real-valued wavenumber. We present formalisms for both TE and TM polarized modes and treatment methods for refractive index jumps. Several numerical examples are considered and the band structures corresponding to different polarizations and incidence angles are obtained.

## 2. FORMALISM

### 2.1 Differential transfer matrix method (DTMM)

Consider a one-dimensional isotropic non-homogeneous medium that is illuminated from a simple ambient medium with refractive index $n_a$ at the angle of $\theta$. This situation is illustrated for layered stratified structures and media with continuous refractive index profiles in Fig. 1. We first consider the analysis of TE polarized waves, and then discuss the extensions to TM polarization. In this case, the wave equation for a non-homogeneous linear medium is given by

$$\frac{\partial^2 A(x)}{\partial x^2} + k^2(x)A(x) = 0, \tag{1}$$

where $A(x)$ is the local amplitude of transverse electric field, and $k(x)$ is the local wavenumber. Here, $k(x) = \sqrt{k_0^2 n^2(x) - \beta^2} \equiv k_0\sqrt{n^2(x) - n_{eff}^2}$, in which $k_0 = \omega/c = 2\pi/\lambda_0$ is the free-space wavenumber with $\omega$, $c$ and $\lambda_0$ being respectively the angular frequency, speed, and wavelength of light in free space. Furthermore, $n(x)$ is the non-homogeneous refractive index function, and $\beta = k_0 n_{eff}$ is the propagation constant. Here $n_{eff} = n_a \sin\theta$ is the propagation effective index, and therefore a constant throughout the calculations. We further assume that either $k(x)$ or $jk(x)$ is real and positive with $j = \sqrt{-1}$. This condition corresponds to lossless media.

According to the DTMM[11], an exact solution of Eq. (1) is

$$A(x) = \exp[-j\mathbf{\Phi}(x)]^t \mathbf{A}(x), \tag{2a}$$

$$\mathbf{A}(x) = \begin{bmatrix} A^+(x) \\ A^-(x) \end{bmatrix}, \tag{2b}$$

$$\mathbf{\Phi}(x) = \begin{bmatrix} +k(x)x \\ -k(x)x \end{bmatrix}, \tag{2c}$$

with $\mathbf{A}(x)$ and $\mathbf{\Phi}(x)$ being the envelope and phase vectors, respectively. Furthermore, $A^+(x)$ and $A^-(x)$ represent the envelope functions associated with the forward and backward waves, and t superscript denotes the transposed matrix. Here, we define the exponential of a non-square matrix to be simply a matrix whose elements are the exponential of corresponding elements of the original matrix.

The envelope vector function $\mathbf{A}(x)$ obeys

$$d\mathbf{A}(x) = \mathbf{U}(x)\mathbf{A}(x)dx, \tag{3a}$$

$$\mathbf{A}(b) = \mathbf{Q}_{a\to b}\mathbf{A}(a), \tag{3b}$$

for any set of two points *a* and *b*, where $\mathbf{Q}_{a \to b}$ is the 2x2 transfer matrix from point *a* to *b*, and is given by

$$\mathbf{Q}_{a \to b} = \exp[\mathbf{M}_{a \to b}] \equiv \exp\left[\int_a^b \mathbf{U}(x)dx\right], \quad (4)$$

and the matrix $\mathbf{U}(x)$ may be shown to be[11]

$$\mathbf{U}(x) = \frac{1}{2k(x)} \frac{dk(x)}{dx} \begin{bmatrix} -1 + j2k(x)x & \exp[+j2k(x)x] \\ \exp[-j2k(x)x] & -1 - j2k(x)x \end{bmatrix}. \quad (5)$$

It is possible to show by direct substitution that Eq. (2) together with Eqs. (3a) and (5) constitutes a solution of Eq. (1). Therefore the formalism is mathematically exact.

The matrix exponential of the square matrix $\mathbf{M}$ is defined as

$$\exp(\mathbf{M}) = \mathbf{I} + \sum_{n=1}^{\infty} \frac{1}{n!} \mathbf{M}^n, \quad (6)$$

where $\mathbf{I}$ is the 2x2 identity matrix. The matrix exponential in Eq. (6) is not to be confused with the exponential operation on a non-square matrix as defined in Eq. (2a). It has the properties $\exp(\mathbf{0}) = \mathbf{I}$, $\exp(\mathbf{M})^{-1} = \exp(-\mathbf{M})$, and $|\exp(\mathbf{M})| = \exp(\text{tr}[\mathbf{M}])$. Moreover, the eigenvalues of $\exp(\mathbf{M})$ are simply given by $\mu_i = \exp(\lambda_i)$ with $\lambda_i$ being the eigenvalues of $\mathbf{M}$.

Therefore, the transfer matrix for any set of three points *a*, *b*, and *c* preserves the properties

$$\mathbf{Q}_{a \to a} = \mathbf{I}, \quad (7a)$$
$$\mathbf{Q}_{a \to c} = \exp(\mathbf{M}_{b \to c} + \mathbf{M}_{a \to b}) = \exp(\mathbf{M}_{b \to c})\exp(\mathbf{M}_{a \to b}) = \mathbf{Q}_{b \to c}\mathbf{Q}_{a \to b}, \quad (7b)$$
$$\mathbf{Q}_{a \to b}^{-1} = \mathbf{Q}_{b \to a}, \quad (7c)$$
$$|\mathbf{Q}_{a \to b}| = \exp(\text{tr}\{\mathbf{M}_{a \to b}\}) = k(a)/k(b). \quad (7d)$$

Note that generally $\mathbf{Q}_{a \to c} = \exp(\mathbf{M}_{a \to b} + \mathbf{M}_{b \to c}) \neq \exp(\mathbf{M}_{a \to b})\exp(\mathbf{M}_{b \to c})$, since the matrices $\mathbf{M}_{a \to b}$ and $\mathbf{M}_{b \to c}$ do not necessarily commute, i.e. $[\mathbf{M}_{a \to b}, \mathbf{M}_{b \to c}] \neq 0$.

### 2.2 Periodic structures
In a periodic system with period L, we have $k(x) = k(x+L)$, and the solution of Eq. (1) according to the Floquet's theorem is given by Bloch waves[1,2,13] as

$$A(x) = \Psi_\kappa(x)\exp(-j\kappa x), \quad (8)$$

in which the envelope function $\Psi_\kappa(x) = \Psi_\kappa(x+L)$ is also periodic. It is possible to show that the dispersion equation for Bloch waves is given by[11]

$$\cos(\kappa L) = \frac{q_{11}}{2}e^{-jk(x)L} + \frac{q_{22}}{2}e^{+jk(x)L}, \quad (9)$$

where $q_{ij}$ are the elements of the transfer matrix $\mathbf{Q}_{x \to x+L}$. The right-hand-side apparently depends on *x* while the left-hand-side does not. However, as shown in our another report[14], the total expression on the right-hand-side of Eq. (9) becomes independent of *x*. Therefore, the choice of *x* on the right-hand-side of Eq. (9) is immaterial.

## 2.3 Even symmetry

Under the even symmetry where $k(x)=k(-x)$, one can take $x=-L/2$ in Eq. (9) so that the elements of $\mathbf{M}_{-L/2 \to L/2}$, $m_{ij}$ simplify to

$$m_{11} = -m_{22} = jL\left[k\left(\tfrac{L}{2}\right) - \bar{k}\right],\tag{10a}$$

$$m_{12} = -m_{21} = j\int_0^{L/2} \frac{\sin[2k(x)x]}{k(x)} k'(x) dx.\tag{10b}$$

Here, the average wavenumber is defined as $\bar{k} = L^{-1}\int_{-L/2}^{L/2} k(x)dx$, and $k'(x) = dk(x)/dx$. Note that since by definition $k(x)$ is proportional to angular frequency $\omega$, so are the diagonal elements $m_{11}$ and $m_{22}$. Therefore in computation of the band-structure, the diagonal elements could be evaluated only once. Using this property of diagonal terms eliminates half of numerical integrations, thus doubling the efficiency. Moreover, if $k(x)$ is a monotonic and hence invertible function of $x$ in the domain $[0,L/2]$, it would be more efficient to change the variables from $x$ to $k$ and evaluate the integral

$$m_{12} = -m_{21} = j\int_{k(0)}^{k(L/2)} \frac{\sin[2x(k)k]}{k} dk.\tag{11}$$

for off-diagonal terms.

## 2.4 Real-valued wavenumber

Now we assume that the wavenumber function $k(x)$ is real-valued over the entire domain $[-L/2, L/2]$. This means that the propagation constant $\beta$ is always smaller than $k_0 n(x)$. This criterion is automatically satisfied when the refractive index of the ambient medium $n_a$ is lower than the minimum refractive index of the periodic medium. It is also satisfied if the angle of incidence is small enough. As a result, at normal incidence the wavenumber function $k(x)$ is real-valued.

It is seen from Eq. (10) that for real-valued $k(x)$, $m_{ij}$ are purely imaginary. Thus $\mathbf{M}_{-L/2 \to L/2}$ has a vanishing trace and under these conditions we have[14]

$$\mathbf{Q}_{-L/2 \to L/2} = (\cosh \lambda)\mathbf{I} + \frac{\sinh \lambda}{\lambda}\mathbf{M}_{-L/2 \to L/2},\tag{12}$$

where $\lambda$ is either one of the eigenvalues of $\mathbf{M}_{-L/2 \to L/2}$.

In Eq. (12), $\mathbf{Q}_{-L/2 \to L/2}$ has the property[14] $q_{11} = q_{22}^*$. Thus, Eq. (9) can be rewritten as

$$\cos(\kappa L) = \text{Re}\{q_{11} \exp[-jk\left(\tfrac{L}{2}\right)L]\},\tag{13}$$

which by using Eqs. (12) and (10a) is simplified into

$$\cos(\kappa L) = \cosh \lambda \cos u + \alpha \sinh \lambda \sin u,\tag{14}$$

with $u = k(L/2)L$ and $\alpha = (u - \bar{k}L)\lambda^{-1}$.

Eq. (14) represents the general dispersion equation of Bloch waves in periodic media with even symmetry and real-valued propagation wavenumbers as discussed above. It is extremely important that the right-hand-side of Eq. (14) is real. For $|\cos(\kappa L)| < 1$, $\kappa$ is real-valued and propagation is within one of the allowed bands. For $|\cos(\kappa L)| = 1$, we have $\kappa L = (2\nu+1)\pi$ with $\nu$ being an integer; this corresponds to propagation at one of the band-gap edges. However, if

$|\cos(\kappa L)|>1$, $\kappa L=(2\nu+1)\pi+j\xi$ is complex with real-valued decay constant $\xi$; this corresponds to propagation within forbidden gaps.

## 2.5 TM polarization

If the propagating wave is TM polarized and the medium is non-magnetic with $\mu=\mu_0$, the corresponding wave equation is

$$\frac{\partial^2 A(x)}{\partial x^2} - 2\frac{\partial \ln[n(x)]}{\partial x}\frac{\partial A(x)}{\partial x} + k^2(x)A(x) = 0, \qquad (15)$$

where $A(x)$ is the local amplitude of transverse magnetic field. As seen above, unlike Eq. (1), the wave equation for magnetic field depends on the derivative of the refractive index profile $n(x)$. However, solutions to (15) can be obtained by DTMM and obey the relations (2)-(4). Therefore, the corresponding transfer matrices should satisfy properties (7). In this case $\mathbf{U}(x)$ in Eq. (5) must be replaced by $\mathbf{V}(x)$ whose elements are given by[11]

$$v_{11}(x) = -\frac{k'(x)}{2k(x)} + \frac{n'(x)}{n(x)} + jk'(x)x. \qquad (16a)$$

$$v_{12}(x) = \left\{\frac{k'(x)}{2k(x)} - \frac{n'(x)}{n(x)}\right\}\exp[+j2k(x)x], \qquad (16b)$$

$$v_{21}(x) = \left\{\frac{k'(x)}{2k(x)} - \frac{n'(x)}{n(x)}\right\}\exp[-j2k(x)x], \qquad (16c)$$

$$v_{22}(x) = -\frac{k'(x)}{2k(x)} + \frac{n'(x)}{n(x)} - jk'(x)x. \qquad (16d)$$

It is interesting to consider on-axis (normal incidence) TM modes for which $\beta=0$ and thus, $k(x)=k_0 n(x)$. In this case $\mathbf{V}(x)$ simplifies to a form similar to $\mathbf{U}(x)$

$$\mathbf{V}(x) = \frac{-1}{2k(x)}\frac{dk(x)}{dx}\begin{bmatrix} -1-j2k(x)x & \exp[+j2k(x)x] \\ \exp[-j2k(x)x] & -1+j2k(x)x \end{bmatrix}. \qquad (17)$$

In any case, Eq. (9) is applicable also to TM modes propagating in a periodic medium with periodicity L. Under even symmetry, Eq. (10a) is still applicable, but Eq. (10b) must be modified appropriately so that the elements $n_{ij}$ of the matrix $\mathbf{N}_{-L/2 \to L/2} = \int_{-L/2}^{L/2} \mathbf{V}(x)dx$ are

$$n_{11} = -n_{22} = jL\left[k\left(\tfrac{L}{2}\right) - \bar{k}\right], \qquad (18a)$$

$$n_{12} = -n_{21} = j\int_0^{L/2} \sin[2k(x)x]\left[\frac{k'(x)}{k(x)} - 2\frac{n'(x)}{n(x)}\right]dx. \qquad (18b)$$

## 2.6 Multi-layer stratified media

Many practical structures are stratified media obtained by periodic stacking of dielectric slabs, as shown in Fig. 1a. In this case, the derivative of $k(x)$ is zero except at finite points coinciding with the interfaces $X_j$, separating the $j$th and $j+1$th layers. In this case, the total transfer matrix of one period $\mathbf{Q}$ can be obtained by multiplying the jump transfer matrices over the interfaces for TE and TM polarized modes, which are given by[7]

$$\mathbf{Q}^{TE}_{j \to j+1} = \begin{bmatrix} \dfrac{k_{j+1}+k_j}{2k_{j+1}}\exp[+j(k_{j+1}-k_j)X_j] & \dfrac{k_{j+1}-k_j}{2k_{j+1}}\exp[+j(k_{j+1}+k_j)X_j] \\ \dfrac{k_{j+1}-k_j}{2k_{j+1}}\exp[-j(k_{j+1}+k_j)X_j] & \dfrac{k_2+k_j}{2k_2}\exp[-j(k_{j+1}-k_j)X_j] \end{bmatrix}, \quad (19a)$$

$$\mathbf{Q}^{TM}_{j \to j+1} = \begin{bmatrix} \dfrac{k_{j+1}+(n_{j+1}/n_j)^2 k_j}{2k_{j+1}}\exp[+j(k_{j+1}-k_j)X_j] & \dfrac{k_{j+1}-(n_{j+1}/n_j)^2 k_j}{2k_{j+1}}\exp[+j(k_{j+1}+k_j)X_j] \\ \dfrac{k_{j+1}-(n_{j+1}/n_j)^2 k_j}{2k_{j+1}}\exp[-j(k_{j+1}+k_j)X_j] & \dfrac{k_2+(n_{j+1}/n_j)^2 k_j}{2k_2}\exp[-j(k_{j+1}-k_j)X_j] \end{bmatrix}, \quad (19b)$$

having determinants $|\mathbf{Q}^{TE}_{j \to j+1}| = k_j/k_{j+1}$, and $|\mathbf{Q}^{TM}_{j \to j+1}| = n_{j+1}^2 k_j / n_j^2 k_{j+1}$, respectively. Here, $n_j$ and $k_j$ are the refractive index and wavenumber in the $j$th layer, respectively. Provided that all $k_j$ are real, all jump transfer matrices $\mathbf{Q}_{j \to j+1}$ across the interfaces obey $q_{11}=q_{22}^*$, and thus so does the overall transfer matrix of one period $\mathbf{Q}_{1 \to n} = \mathbf{Q}_{n-1 \to n}\mathbf{Q}_{n-2 \to n-1}...\mathbf{Q}_{1 \to 2}$, with $n$ being the total number of interfaces or layers in a period. Therefore, the Eq. (13) is still applicable in the form

$$\cos(\kappa L) = \mathrm{Re}\{q_{11}\exp[-jk_1 L]\}, \quad (20)$$

with $q_{11}$ being the first element of $\mathbf{Q}_{1 \to n}$. Here, the first layer is characterized as the leftmost layer in the selected period window.

## 3. APPLICATION EXAMPLES

In this section, we discuss a few application examples for the computation of band structures of periodic media both with and without even symmetry. Here, three symmetric and one asymmetric profile are used as shown in Fig. 2: sinusoidal, triangular, piece-wise step, and ramp with jump. All profiles are considered to have a large contrast, varying between 1 and 3. It is also assumed that the periodic medium is illuminated from air with $n_a=1$, at the angles of incidence 0 and $\pi/4$.

Band-structures are separately shown for TE and TM modes. In Fig. 3, the first forbidden gap is shown. Figures 3a, 3b, 3c, and 3d respectively, correspond to sinusoidal, triangular, piece-wise step, and asymmetric profiles. Figure 3 shows that the band structures for all profiles at low frequencies are linear. This is exactly what expected from general behavior of a periodic medium, whose propagation characteristics at long wavelengths are given by proportional frequency and wavenumber, i.e. identical phase and group velocities. However, as frequency is increased the system deviates from the low frequency behavior and group velocity reduces gradually to zero at $\kappa L=\pi$. At this point, increasing frequency makes the wavenumber complex and thus the propagating modes is evanescent. Here, we have entered the first forbidden gap. The size of the first energy gap is dependent on the polarization, angle of incidence, as well as the index profile.

At the upper frequency edge of the first forbidden gap, the group velocity is again zero. As we further increase the frequency, the group velocity gradually increases to a maximum and then decreases to zero just before entering the second forbidden gap. It can be seen from Fig. 3 that forbidden gaps for TM modes are larger than those for TE modes with the same profile and incidence angle. Additionally, band gaps produced by step-wise and triangular profiles are the largest and smallest, respectively, in symmetric profiles. Meanwhile, the asymmetric profile with a jump produces very large second gap compared to the symmetric structures.

Finally, it is remarkable to note that band structures of TE and TM modes at normal incidence are exactly the same, both for symmetric and asymmetric profiles, in agreement to what claimed earlier. The total computation time for each curve, for a high-level MATLAB® code running on a Windows platform has been typically a few seconds. The computation time with the same accuracy using plane-wave expansion method would be roughly an order of magnitude higher on the same computer and language platform.

## 4. CONCLUSIONS

We presented here the application of differential transfer matrix method (DTMM) for finding the exact dispersion equation for wave propagation in one-dimensional periodic media. The problem was greatly simplified by the assumption of even symmetry and real wavenumber function resulting into a simple governing equation. We discussed the calculation of band structure for both TE and TM modes at different incidence angles. We also derived formulae for the calculation of band structure in stratified layered media. Finally, several examples have been presented to show the applicability of the approach.

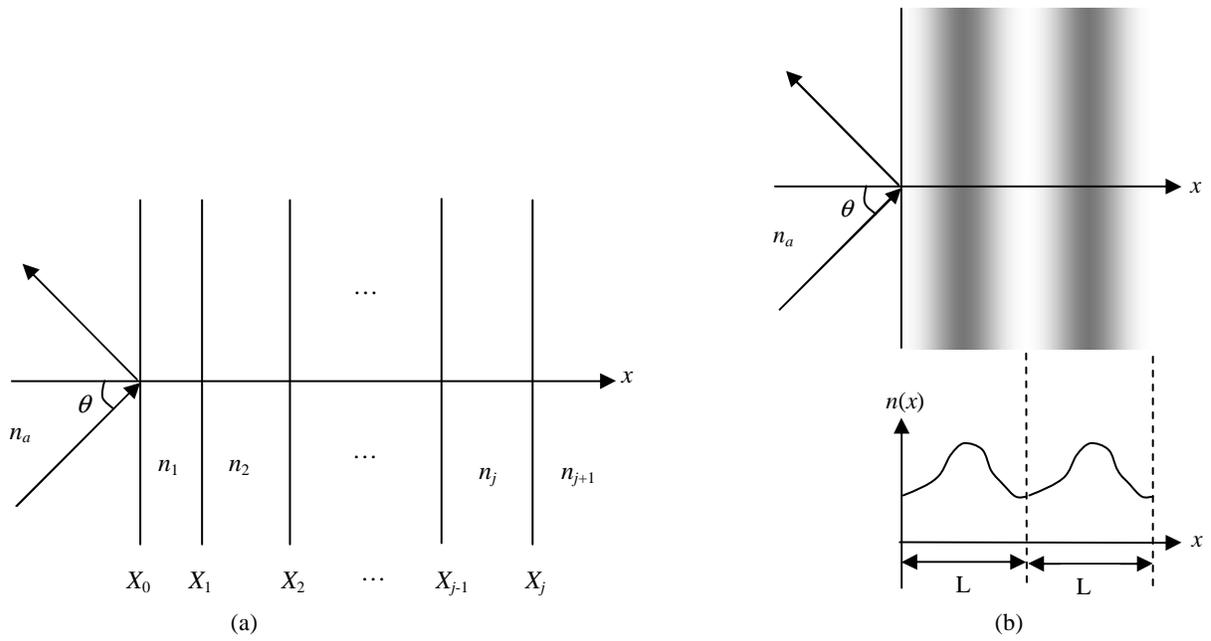

Fig. 1. Illustration of a non-homogeneous structure illuminated from a simple ambient medium:
(a) layered stratified structure; (b) continuous refractive index profile.

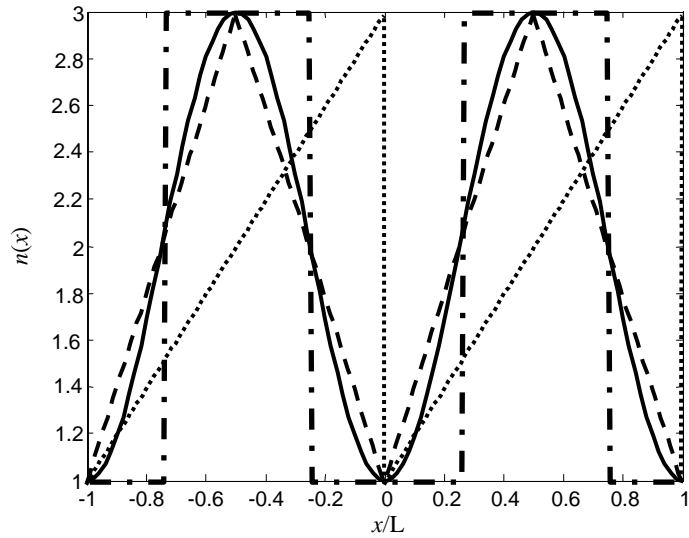

Fig. 2. Three basic symmetric periodic refractive index profiles:
(−) sinusoidal; (--) triangular; (−·) square, and an asymmetric periodic refractive index profile with a jump (··).

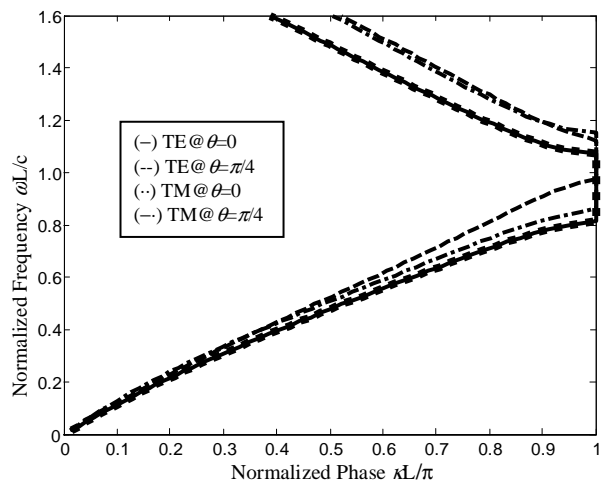

(a)

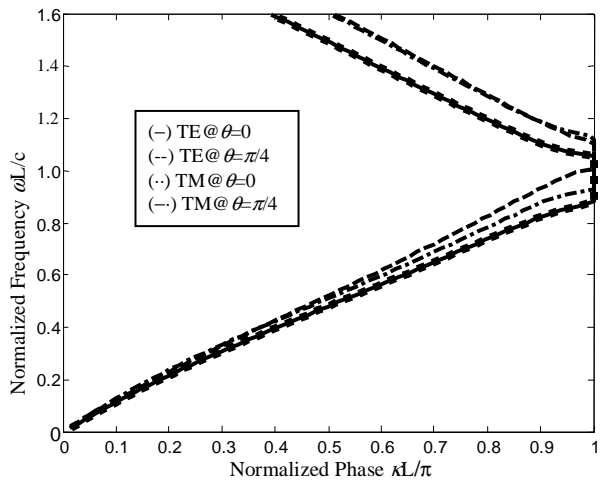

(b)

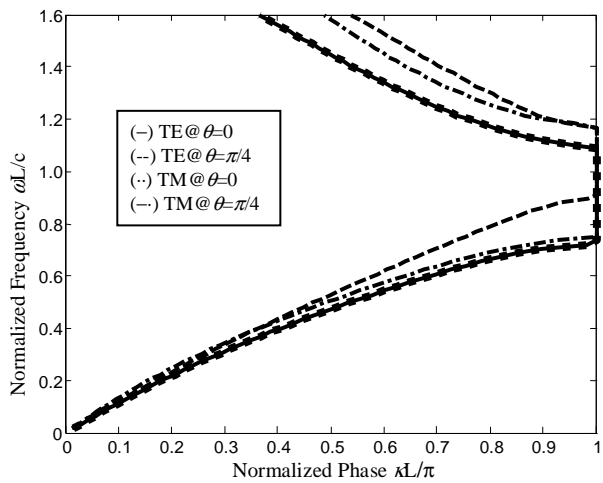

(c)

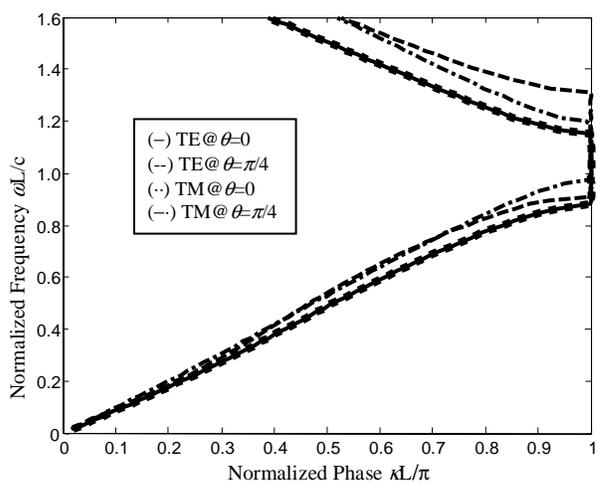

(d)

Fig. 3. Band structure of different polarizations and incident angles around the first band gap
for periodic media with index patterns shown in Fig. 2:
(a) sinusoidal; (b) triangular; (c) square; (d) asymmetric.